\documentclass[aps,pra,twocolumn,superscriptaddress]{revtex4}

\usepackage[dvips]{graphicx}
\usepackage{amsmath}
\usepackage{amsfonts}
\usepackage{amssymb}
\usepackage[latin1]{inputenc}
\usepackage{xcolor}

\definecolor{darkblue}{rgb}{0, 0, 0.8}
\usepackage[colorlinks=true, breaklinks=true, linkcolor=darkblue, citecolor=darkblue, urlcolor=darkblue]{hyperref}
\newcommand{\doilink}[2]{\href{http://dx.doi.org/#1}{#2}}

\newcommand{\beq}{\begin{equation}}
\newcommand{\eeq}{\end{equation}}

\begin{document}
\title{Homogenization of an ensemble of interacting resonant scatterers}
\date{\today}
\author{N.J. Schilder} 
\affiliation{Laboratoire Charles Fabry, Institut d'Optique Graduate School, CNRS, Universit\'e Paris-Saclay, 91127 Palaiseau Cedex, France}
\author{C. Sauvan}
\affiliation{Laboratoire Charles Fabry, Institut d'Optique Graduate School, CNRS, Universit\'e Paris-Saclay, 91127 Palaiseau Cedex, France}
\author{Y.R.P.~Sortais}
\affiliation{Laboratoire Charles Fabry, Institut d'Optique Graduate School, CNRS, Universit\'e Paris-Saclay, 91127 Palaiseau Cedex, France}
\author{A.~Browaeys} 
\affiliation{Laboratoire Charles Fabry, Institut d'Optique Graduate School, CNRS, Universit\'e Paris-Saclay, 91127 Palaiseau Cedex, France}
\author{J.-J.~Greffet}
\affiliation{Laboratoire Charles Fabry, Institut d'Optique Graduate School, CNRS, Universit\'e Paris-Saclay, 91127 Palaiseau Cedex, France}

\begin{abstract}
We study theoretically the concept of 
homogenization in optics using an ensemble of randomly distributed 
resonant  stationary atoms with density $\rho$. 
The ensemble is dense enough for the usual condition for 
homogenization, viz. $\rho\lambda^3 \gg 1$, to be reached. 
Introducing the coherent and incoherent scattered powers, we define 
two criteria to define the homogenization regime. 
We find that when the excitation field is tuned in a broad frequency range
around the resonance, 
none of the criteria for homogenization is fulfilled, meaning
that the condition $\rho\lambda^3\gg 1$ is not sufficient to 
characterize the homogenized regime around the atomic resonance. We interpret 
these results as a consequence of the light-induced dipole-dipole interactions 
between the atoms, which implies a description of scattering in terms of 
collective modes 
rather than as a sequence of individual scattering events. 
Finally, we show that, although 
homogenization can never be reached for a dense
ensemble of randomly positioned laser-cooled atoms around resonance, 
it becomes possible if one introduces 
spatial correlations in the positions of the atoms 
or non-radiative losses, such as would be the case for organic molecules 
or quantum dots coupled to a phonon bath.
\end{abstract}

\maketitle

\section{Introduction}

Homogenization is the procedure by which one replaces a 
discrete distribution of  particles by a continuous density distribution. 
In the framework of the electrodynamics of continuous media, 
the standard procedure of homogenization, described in many textbooks~\cite{Jackson,Ashcroft,Zangwill2012}, 
supposes that the inter-particle distance $\rho^{-{1\over 3}}$ 
($\rho$ is the spatial density) is much smaller 
than the characteristic length scale associated with the phenomenon 
under study, usually the propagation of a wave through the medium. 
The characteristic scale being the wavelength  $\lambda$, 
the condition for homogenization is thus assumed to be $\rho\lambda^3\gg1$. 
When this condition is satisfied, 
one derives the \textit{macroscopic} properties 
of an ensemble of scatterers from 
the \textit{microscopic} properties of each of them
by means of an effective medium theory. In the context of optics, for example, 
several models exist that relate the (microscopic) atomic 
polarizability~\cite{Frisch1968} or the dielectric constants 
of spherical nano-particles in a composite dielectric random medium~
\cite{Maxwell1904,Bruggeman1935,Sheng1995,Lagendijk1996,Greffet2007,Sentenac2005} to the 
(macroscopic or effective) dielectric constant of the system. 

In order to derive criteria for homogenization in optics, 
one usually decomposes the electric field scattered by an ensemble of 
scatterers (e.g. atoms)
into coherent and incoherent (or diffuse) components, 
$\langle\textbf{E}_{\rm sc}\rangle$ 
and $\delta{\bf E}_{\rm sc}$ respectively: 
$\textbf{E}_{\rm sc}=\langle\textbf{E}_{\rm sc}\rangle+\delta\textbf{E}_{\rm sc}$, 
where $\langle\delta\textbf{E}_{\rm sc}\rangle=0$. Here $\langle.\rangle$ denotes an 
ensemble average over many different 
spatial realizations.  
The {\it coherent} (or average) monochromatic field 
$\langle\textbf{E}_{\rm sc}\rangle$ follows 
the Helmholtz equation with an  effective (i.e. ensemble-averaged)
dielectric constant 
$\epsilon_{\rm eff}(\omega,{\bf r})$ 
describing the medium~\cite{Frisch1968,Sheng1995,Lagendijk1996,Greffet2007}: 
\beq
\nabla\times\nabla\times\langle{\bf E}_{\rm sc}\rangle-\epsilon_{\rm eff}(\omega,{\bf r})
\frac{\omega^2}{c^2}\langle{\bf E}_{\rm sc}\rangle=0\ .
\eeq 
Importantly one can always associate
an effective dielectric constant (and therefore an effective index of refraction) 
to this coherent 
component, no matter whether the medium is homogeneous or not, and even in the presence
of a strong incoherent field.  
The  {\it incoherent} 
scattered field $\delta\textbf{E}_{\rm sc}$ originates 
from the random positions of the scatterers in the ensemble. 
These two components lead to the coherent and incoherent scattered powers,   
$P_{\rm coh}$ and $P_{\rm incoh}$. 
Daily life experience indicates that 
a gas of atoms or molecules, such as the atmosphere, scatters light efficiently
away from the direction of the incoming light beam. 
However, most of the scattered power is coherent and  in the direction of the incoming  beam. 
This suggests a first, and weak criterion for homogenization:
$P_{\rm incoh}/P_{\rm coh}\rightarrow 0$ when 
the density of scatterers increases. In this case, the question 
is therefore whether the effective dielectric constant is enough to describe accurately 
the propagation of light in the ensemble of scatterers, since coherent light 
scattering dominates. A second,  stronger criterion
comes from the observation that pure water or an amorphous glass, which are 
dense materials with $\rho\lambda^3\gg1$,
do not scatter light:  there is therefore 
no incoherent scattering ($P_{\rm incoh}= 0$) in this homogenized situation. 
From these examples,
it would seem that the condition $\rho\lambda^3\gg 1$ leads to homogenization
according to at least one of the two criteria described above.

In this work, however, we show theoretically that this condition $\rho\lambda^3\gg 1$ 
is not sufficient in the   
case of dense ensembles of randomly positioned {\it resonant} scatterers. 
Our study is 
motivated by recent experimental developments, which now make it possible to 
prepare ensembles of resonant scatterers in volumes comparable or 
smaller than the wavelength  of an optical transition, 
such as ensembles of quantum dots~\cite{Novotny2006,Belacel2013} 
or clouds of laser-cooled atoms~
\cite{Pellegrino2014,Jennewein2016}. In particular we have recently measured both the 
incoherent~\cite{Pellegrino2014} and coherent~\cite{Jennewein2016} response 
of a wavelength-sized cold atomic cloud for which the atomic density can be varied.
The main result of the present work is the 
finding that this ensemble of atoms submitted to a near-resonant light field can {\it never} 
be homogenized according to both criteria presented above. 
The situation is all the more striking that, as we have shown in a previous work~\cite{Nick2016}, 
the effective index of refraction of the cloud can be as large as 2, a value even larger than for 
many condensed matter systems for which homogenization has been extensively proven to work. 
We show that the origin of this feature lies in the light-induced dipolar interactions between the 
atoms that are very strong when the light is tuned near-resonance. 
They lead to a collective response of the medium that has to be described in terms of modes
rather than in term of individual atoms.  

The paper is organized as follows. We first detail the model 
used to calculate the coherent and incoherent scattered powers. 
In Sec.~\ref{Sec:off_resonance}, we apply our formalism to the far-off resonance 
case,  and check that the weak criterion for homogenization applies.
Section~\ref{Sec:failureHomo} 
presents the transition from the far-off to the 
near-resonant case and the fact that a dense cloud of laser-cooled atoms can never 
be homogenized in the latter regime according to both criteria. 
We then give a first interpretation in terms of
collective modes (Sec.~\ref{Sec:Interpret_modes}) and a second one, which  considers
the cloud as an effective medium described by a
dielectric constant
(Sec.~\ref{Sec:interpret_dielectric}). 
In Sec.~\ref{Sec:correlatio_positions}, we discuss the influence of
correlations in the positions of the scatterers and observe that they lead to a recovery of 
homogenization according to the two criteria. 
Finally (Sec.\ref{Sec:NRlosses}), 
we show that introducing a non-radiative decay 
channel also leads to a recovery of homogenization. 

\section{Description of the system}\label{Description}

\begin{figure}
\includegraphics[width=1\linewidth]{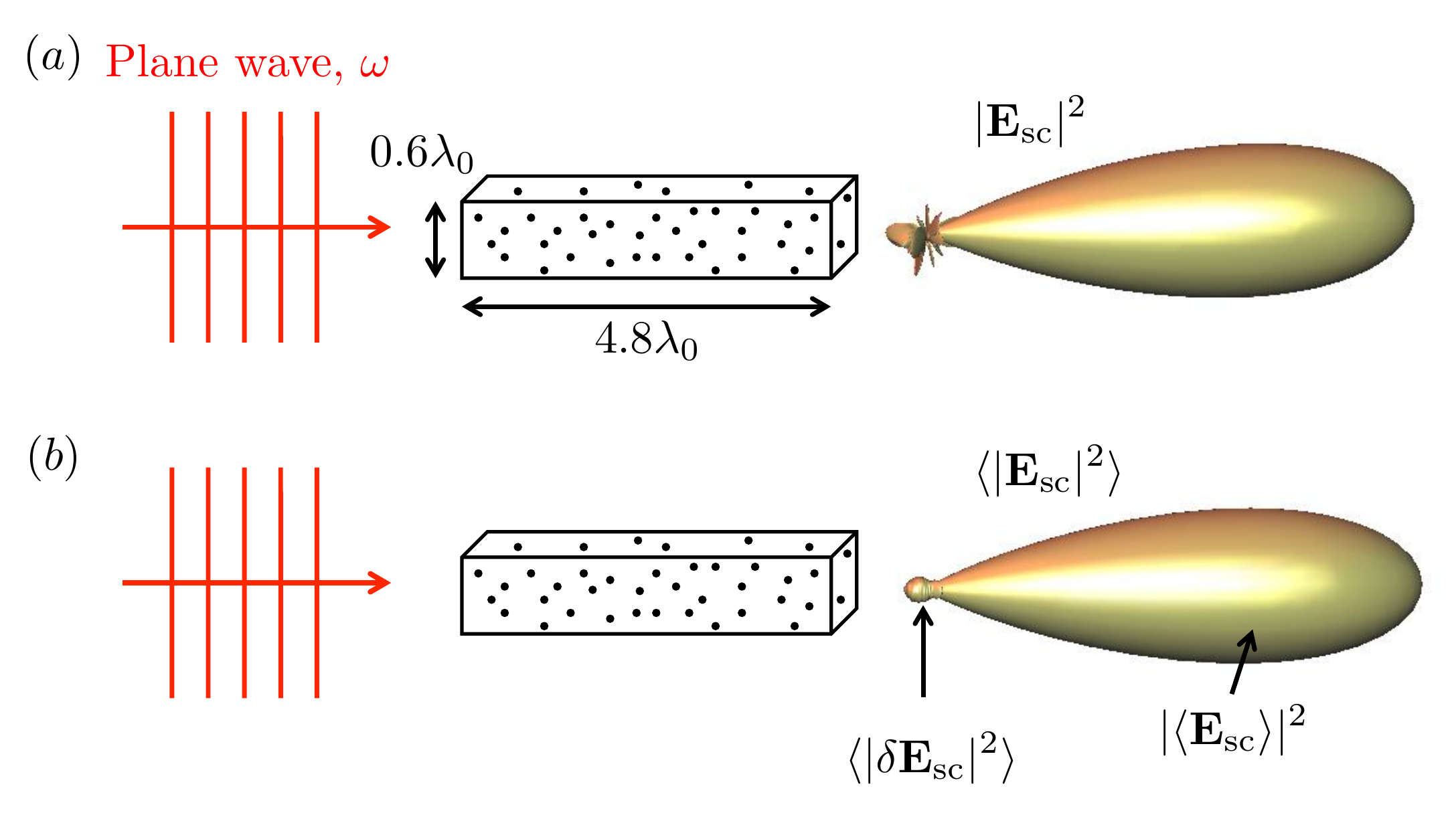}
\caption{(a) Scattering pattern $|{\bf E}_{\rm sc}|^2$
for a single realization of a cloud of 450 atoms
(volume $4.8\lambda_0\times0.6\lambda_0\times0.6\lambda_0$)
illuminated by a plane wave on resonance with the atomic transition
($\omega=\omega_0$).  (b) Scattering pattern averaged over 100 realizations
of the distribution of atomic positions ($N=450$, $\omega=\omega_0$).
The speckle structure associated to the incoherent scattering 
$\langle|\delta{\bf E}_{\rm sc}|^2\rangle$ does not show a preferred direction,
while the forward direction is dominated by the coherent 
scatterering $|\langle{\bf E}_{\rm sc}\rangle|^2$.}
\label{Fig1}
\end{figure}

We study theoretically the scattering of a plane wave 
(frequency $\omega$, wavevector $k=\omega/c=2\pi/\lambda$) from a disordered, 
wavelength-size cloud of cold atoms, 
as illustrated in Fig.~\ref{Fig1}. We idealize the experimental situation 
of our previous works~\cite{Pellegrino2014,Jennewein2016} 
by assimilating the cloud to a parallelepiped with dimensions 
$V=4.8\lambda_0\times0.6\lambda_0\times0.6\lambda_0$, where $\lambda_0=780$ nm 
is the resonance wavelength of the D2 transition of rubidium-87 atoms 
used in the experiment. 
The atoms are uniformly distributed (spatial density $\rho = N/V$) and modeled 
as point-like and 
identical scatterers characterized by an isotropic  polarizability:
\beq
\alpha(\omega)=\frac{3\pi\Gamma_0/k^3}{\omega_0-\omega-i\frac{\Gamma_0+\Gamma_{\text{nr}}}{2}},
\eeq
with $\omega_0 =2\pi c/\lambda_0$ the transition frequency, $c$ the speed of 
light in vacuum, $\Gamma_0$ and $\Gamma_{\text{nr}}$ respectively the 
radiative and non-radiative decay rate ($\Gamma_0=2\pi\times 6\text{ MHz}$ for Rb). 
When $\Gamma_{\text{nr}}=0$, this polarizability model corresponds to a 
classical $J=0\rightarrow J=1$ atom, where $J$ 
is the angular momentum~\footnote{Here we neglect the complex internal structure of the rubidium
atom used in the experiment.}. This model can also include 
non-radiative decay channels ($\Gamma_{\text{nr}}$), as would be 
necessary if we were discussing e.g., systems of
quantum dots. 
However, unless stated differently, we assume $\Gamma_{\text{nr}}=0$, which is a 
good model for a cold atomic gas. Finally, the scattering cross section is
given by $\sigma_{\rm sc}(\omega)=k^4|\alpha(\omega)|^2/(6\pi)$~\cite{Jackson}.

As we discuss dense atomic systems, i.e. $\rho/k^3\gtrsim 1$, 
we include the resonant dipole-dipole interactions 
between the atoms. As explained, e.g. 
in Refs.~\cite{Pellegrino2014,Jennewein2016,Nick2016}, 
the dipole of atom
$j$ is driven by the laser field and the field radiated by all the other atoms.  
This approach leads to a set of coupled dipole
equations, in steady-state: 
${\bf p}_j=\epsilon_0 \alpha ({\bf E}_{{\rm L}j}+\mu_0\omega^2\sum_{i\neq j} [{\bf G({\bf r}_i-{\bf r}_j)}]{\bf p}_i)$, 
with ${\bf E}_{{\rm L}j}$ the field of the laser at the position of atom $j$. 
Here, the Green's tensor $[{\bf G({\bf r}_i-{\bf r}_j)}]$ describes the resonant dipole-dipole 
interactions between atoms $i$ and $j$ including  
the $1/r$, $1/r^2$ and $1/r^3$ terms. 
After solving the set of equations to get each dipole moment, 
we calculate the scattered electric field 
$\textbf{E}_\text{sc}({\bf r})=\mu_0\omega^2\sum_{i}[{\bf G({\bf r}-{\bf r}_i)}]{\bf p}_i$ (for more details, see~
\cite{Pellegrino2014,Jennewein2016,Nick2016}). 

After few hundreds of realizations for which the 
atomic positions are changed according to a 
uniform probability distribution, 
we calculate the scattered field $\textbf{E}_{\rm sc}$, 
the coherent, ensemble-average field
$\langle\textbf{E}_{\rm sc}\rangle$ and the incoherent, fluctuating field 
$\delta\textbf{E}_{\rm sc}=\textbf{E}_{\rm sc}-\langle\textbf{E}_{\rm sc}\rangle$.
The ensemble-averaged scattering pattern is then 
decomposed in its coherent and incoherent parts: 
$\langle|\textbf{E}_\text{sc}|^2\rangle=|\langle\textbf{E}_\text{sc}\rangle|^2+\langle|\delta\textbf{E}_\text{sc}|^2\rangle$. 
The coherent scattering pattern corresponds to the diffraction 
pattern of a homogeneous 
object described by an effective dielectric constant and the incoherent 
scattering pattern is a quasi-isotropic speckle originating from the 
random positions of the atoms in the cloud (see Fig.~\ref{Fig1}).  

In order to characterize quantitatively the level of homogenization, 
we define the integrated scattered powers corresponding respectively 
to the ensemble-averaged 
fields $|\langle\textbf{E}_\text{sc}\rangle|^2$ and  
$\langle|\delta\textbf{E}_\text{sc}|^2\rangle$ evaluated on a 
spherical surface $\Sigma$ in 
the far field: 
\beq\label{Eq:Pcoh}
P_{\text{coh}}=\frac{\epsilon_0c}{2}\oint_{\Sigma} |\langle\textbf{E}_{\text{sc}}\rangle|^2dS
\eeq
and 
\beq\label{Eq:Pinc}
P_{\text{incoh}}=\frac{\epsilon_0c}{2}\oint_{\Sigma} \langle|\delta\textbf{E}_{\text{sc}}|^2\rangle dS.
\eeq
The total scattered power is then 
$P_{\text{tot}}=P_{\text{coh}}+P_{\text{incoh}}$. It does not include the incident field. 

\section{Far-off resonance light scattering}\label{Sec:off_resonance}

\begin{figure}
\begin{centering}
\includegraphics[width=\linewidth]{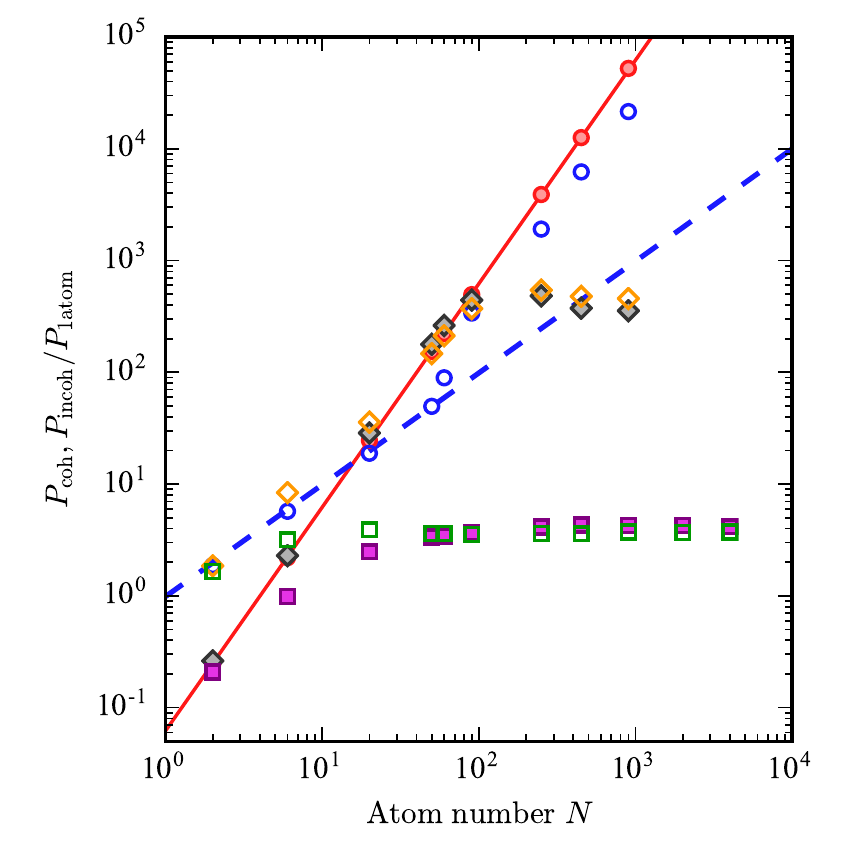}
\caption{Coherent (plain symbols)  and incoherent (open symbols) 
scattered powers 
calculated for a linearly polarized 
incident plane wave propagating along the long axis of the 
cloud for various frequency detunings 
$\Delta=\omega-\omega_0$, as a function 
of the number of atoms $N$. 
All powers are normalized to the power scattered by 
a single atom at the same detuning. 
The plain (dashed) line connects the values of the coherent (incoherent) power 
for $\Delta=-10^4 \Gamma_0$. Circles:  $\Delta=-500 \Gamma_0$.
Diamonds: $\Delta=-5 \Gamma_0$. Squares: $\Delta=0$.
}
\label{Fig:faroff_to_near_resonance}
\end{centering}
\end{figure}

Figure~\ref{Fig:faroff_to_near_resonance} presents the coherent and incoherent scattered
powers as a function of the number of atoms $N$ inside the fixed volume of the cloud,
normalized by the scattered power of a single atom, for different 
detunings $\Delta=\omega-\omega_0$ of the incoming plane 
wave with respect to the atomic transition. This figure 
allows us to explore the transition 
between far-off to near-resonance scattering. 

When the laser is very far-off resonance 
($\Delta=-10^4\Gamma_0$), 
we observe the scalings  $P_{\text{incoh}}\propto N$, and $P_{\text{coh}}\propto N^2$.
As reminded in  Appendix~\ref{Appendix:single_scattering},
this result is expected as the wavelength-size atomic cloud 
is in the single scattering regime~\cite{Berkeley}: 
the mean free path $\ell_\text{sc}=1/[\rho\sigma_\text{sc}(\omega)]=3\text{ m}$ for $N=450$ 
atoms is much larger than the size of the atomic cloud. In this regime of large detuning, 
the weak criterion for homogenization applies, as 
$P_{\rm incoh}/P_{\rm coh}\propto 1/N\rightarrow 0$ when $N$ increases. 
We note from  Fig.~\ref{Fig:faroff_to_near_resonance} that the 
cloud enters the homogenization 
regime for $N\gtrsim20$, i.e., $\rho\lambda^3>1$. 

The fact that, in the single scattering limit, $P_{\rm incoh}\propto N$ is extensively 
used to calibrate the number of atoms in a cloud of e.g. cold atoms in 
experiments on laser cooling or quantum degenerate gases~\cite{Ketterle_Varenna_1998}.
It is also common in these experiments to measure the index of refraction
of the atomic sample by measuring the coherent optical response of the cloud
using e.g. absorption or phase contrast imaging.
This again emphasizes the fact that one can always define an 
index of refraction (or a dielectric constant) to characterize the coherent response
of the cloud, even in the presence of incoherent scattering.  

\section{From far-off to near-resonance scattering: failure of homogenization}\label{Sec:failureHomo}

Coming back to Fig.~\ref{Fig:faroff_to_near_resonance}, we observe that 
when the detuning gets closer to resonance, the scaling laws for both
the coherent and incoherent scattering as a function of the atom number  
are strongly modified with respect to the far-off resonance case. 

When  $\Delta=-500\Gamma_0$, the coherent power still follows 
$P_{\rm coh}\propto N^2$. However, and quite unexpectedly, the incoherent power
also features the same $N^2$ dependence for $N\gtrsim 30$, despite the fact that 
the cloud still operates in the single-scattering limit 
($\ell_\text{sc}=3$ mm for $N=1000$). When the detuning
is close to resonance ($\Delta=-5\Gamma_0$), the coherent and incoherent powers are
nearly identical for $N\gtrsim 30$, and saturate when $N\gtrsim 100$. 
Finally, at resonance ($\Delta=0$), 
in stark contrast with the off-resonance scattering case, 
the two powers
become rapidly independent of $N$ and
saturate to the approximately same value when the number of atoms increases. 
This saturation of both the incoherent and coherent scattered powers was actually observed 
in our recent experimental works~\cite{Pellegrino2014,Jennewein2016}, 
although there we could only
measure a fraction of the powers in a given solid angle. 
As a consequence, as far as the homogenization of the ensemble of atoms is concerned,
neither the weak ($P_{\rm incoh}/P_{\rm coh}\rightarrow 0$) 
nor the strong criterion ($P_{\rm incoh}=0$) apply for detunings in the range 
$|\Delta/\Gamma_0|\lesssim 500$, although the condition
$\rho\lambda^3\gg1$ is fulfilled.
It thus appears that {\it a dense cloud of cold atoms can never be homogenized 
in a broad frequency range  around the resonance}! 
In other words: a dense cloud of cold atoms keeps scattering a lot of incoherent
light, and homogenization is reached very 
far from resonance, and only according to the weak criterion.

Before we give a consistent interpretation of this fact in the next sections, 
we further explore the resonant case. Figure \ref{Fig1}(b) 
shows the ensemble-averaged scattering pattern of the cloud. 
We observe that the amplitude of 
the speckle is very low in the forward direction compared 
to the coherent scattering. 
In other directions,  incoherent scattering dominates. 
From the figure, the coherent scattered power seems 
to dominate the total power. However as it is scattered  
in a limited solid angle
as opposed to the incoherent power, 
after integration 
over all directions it turns out  that  the coherent and incoherent powers have 
similar values in our case (see Fig.~\ref{Fig:faroff_to_near_resonance}). 

\begin{figure}
\includegraphics[width=\linewidth]{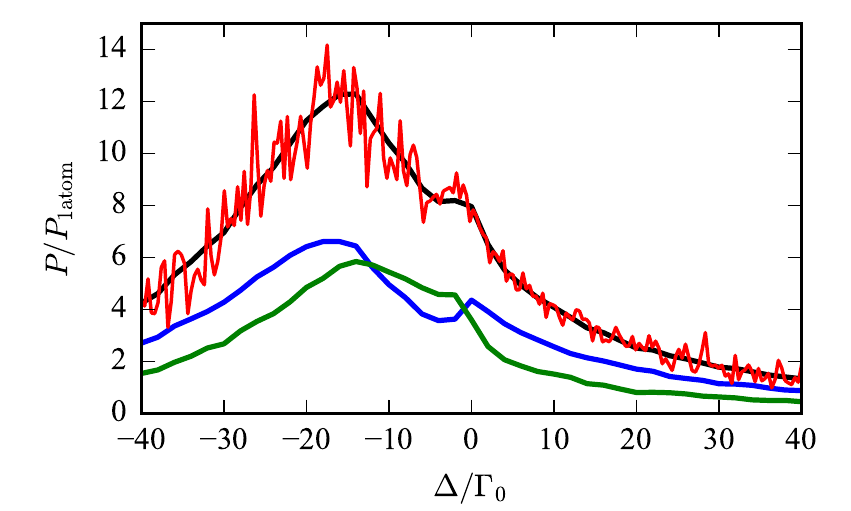}
\caption{Total scattered power for 
a single realization (red) of a cloud of $N=450$ atoms, 
together with the ensemble 
averaged coherent (blue), incoherent (green) and total (black) scattered powers. 
All quantities are normalized to the power scattered by a single atom at resonance.}
\label{Fig:spectrum_resonance}
\end{figure}

Finally, we calculate the powers scattered coherently and incoherently 
as a function of the detuning of the laser, near the atomic resonance. 
We plot in Fig.~\ref{Fig:spectrum_resonance} 
four curves corresponding to
(1) the ensemble-averaged coherent scattered light, 
(2) the ensemble-averaged incoherent scattered light, 
(3) the sum of the ensemble-averaged coherent and incoherent 
scattered light, and 
(4) the scattered light for a single realization of the distribution of atoms 
in the cloud. Firstly, we observe that the line shapes are significantly 
different from a Lorentzian, contrarily to what it would be in the single scattering regime. 
Secondly, the coherent and incoherent powers
exhibit similar shapes, in particular a double structure with a peak for a 
negative detuning. 

All the features presented in this section indicate that the interpretation 
of scattering as a sequence of individual scattering events breaks down 
in a broad frequency range around resonance,
despite the fact that for all the atom numbers used in this work, 
the mean-free path $\ell_\text{sc}$ 
is always at least 10 times larger than the
cloud largest size, even on resonance. We would otherwise observe a lorentzian
line shape only. This emphasizes that the definition of the  mean free path 
by $\ell_\text{sc}=1/[\rho\sigma_\text{sc}(\omega)]$ is not appropriate
to describe the scattering of light in our situation. Instead 
the correct length scale is associated with the decay of the field in the medium 
and is given by $1/(n'' k)$, with $n''$ the imaginary part of the effective refractive index. As 
$n''$ reaches values as high as 2~\cite{Nick2016}, the length scale is 
100 nm, smaller than the size of the cloud. 

In the next two sections
we interpret the above observations using two different, but complementary 
points of view.  

\section{Interpretation in terms of collective modes}\label{Sec:Interpret_modes}

As discussed by many authors (for recent works, 
see e.g.~\cite{Ruostekoski1997,Scully2010,Chomaz2012,Kupriyanov2013,Zhu2013,Kaiser2014,Bettles2015,Bettles2016,Nick2016}), the coupling 
between atoms resulting from the light-induced dipole-dipole interaction 
leads to a set of collective modes $\beta$ of the atomic dipoles. These modes
are eigenstates of the set of coupled dipole equations. 
Each of these $3N$, non-orthogonal modes
has its own (complex) eigenfrequency $\tilde{\omega}_\beta=\omega_0+\Omega_\beta
-i\frac{\Gamma_\beta}{2}$, with $\Omega_\beta$ the collective shift and $\Gamma_\beta$
the collective decay rate. Some of these modes, featuring $\Gamma_\beta<\Gamma_0$,
are sub-radiant, while others with $\Gamma_\beta>\Gamma_0$ are super-radiant. 
In Ref.~\cite{Nick2016}, we studied in detail the modes corresponding
to the situation analyzed here and depicted in Fig.~\ref{Fig1}. We found that the
modes fall in three categories (see Fig. 2 of Ref.~\cite{Nick2016}). The first one
consists of modes with collective linewidth $\Gamma_\beta\approx 2\Gamma_0$ 
(super-radiant) and $\Gamma_\beta \ll \Gamma_0$ (sub-radiant). These so-called 
dimer modes
are made of pairs of atoms and have large 
collective frequency shifts $\Omega_\beta \sim {\Gamma_0/(kr)^3}$, with $r$ the interatomic
distance. The second category consists of few polaritonic 
modes that have four key features: 
(i) all atoms have significant contributions to the modes
(ii) they are robust against disorder (i.e., they depend on density and 
geometry but not on the precise positions of the scatterers), 
(iii) they are super-radiant with $\Gamma_\beta \gtrsim10 \Gamma_0$ and 
(iv) although representing less than 
1\% of the total number of modes, we calculated that
they contribute for a large ($>50\%$) fraction 
of the coherent 
scattering of the cloud.  
Finally, the last category consists of modes with the excitation delocalized
over many atoms, but with no regular spatial structure.  They can be super- or 
sub-radiant and their frequency shift is on the order of a few $\Gamma_0$.

We can now interpret 
the transition between far-off to near-resonance 
scattering observed in Fig.~\ref{Fig:faroff_to_near_resonance}
in terms 
of collective modes. First, when the incoming light
is very far-detuned, with $|\Delta|$ larger than the largest shift corresponding to the 
pair of closest atoms in the cloud (see Fig. 2 of Ref.~\cite{Nick2016}), 
the light cannot excite any mode.  
In this regime, the cloud scatters the light as a collection of independent atoms, with 
the usual scaling laws $P_{\text{incoh}}\propto N$ and $P_{\text{coh}}\propto N^2$.

When the detuning  becomes small enough that dimer modes 
are excited (see Fig. 2 of Ref.~\cite{Nick2016}), the 
incoherent scattering  is dominated by the scattering from the super-radiant 
pairs of atoms, which is 
essentially isotropic for atoms closer than $1/k$. The incoherent
scattered power is therefore proportional to the number of excited pairs, 
hence $\propto N^2$. As a consequence,
a description in terms of dimer modes naturally 
leads to a $N^2$ scaling for the incoherent 
scattered power. 
A semi-classical model, presented in  Appendix~\ref{Appendix:dimer_modes},
predicts  a critical number of atoms $N_{\rm c}\approx 25$ 
where the transition between scattering by individual atoms 
to scattering by dimers occurs and is in good agreement 
with the value observed  
in Fig.~\ref{Fig:faroff_to_near_resonance}. 
As for the coherent power, we show in Appendix~\ref{Appendix:dimer_modes}
that  it is still dominated by the coherent scattering from individual atoms, with a negligible 
contribution from the dimers. Therefore, the coherent power varies as $N^2$ 
as in the far-off resonance case 
(see Sec.~\ref{Sec:off_resonance} and Appendix~\ref{Appendix:single_scattering}). 

Finally, when the detuning becomes very close to resonance, many 
delocalized (non-polaritonic) modes are excited. 
Incoherent scattering comes from many
of these modes, which have no regular spatial structure. 
In our intermediate regime $V\sim \lambda^3$, we could not derive 
any simple scaling law for the incoherent scattering as a function 
of the atom number. 
However, qualitatively, the saturation of the {\it incoherent} power observed 
in Fig.~\ref{Fig:faroff_to_near_resonance} 
indicates that, when
the number of atoms increases beyond $\rho /k^3\gtrsim 1$,
the product of the number of modes excited in 
a bandwidth of the order of $\Gamma_0$ by
the power they radiate has to 
become independent of the number of atoms. 
As for the {\it coherent} scattered power, 
few polaritonic modes have important contributions. 
Their collective shifts and widths are mainly set by the  geometry of the cloud
and hardly depend on the number of atoms~\cite{Nick2016}, leading to a saturation of the
coherent power. Finally, we have checked that the saturation of the coherent and incoherent
powers at comparable values (see Fig.~\ref{Fig:faroff_to_near_resonance}) 
is a consequence of the particular 
dimensions of cloud.   

\section{Interpretation in terms of effective medium}\label{Sec:interpret_dielectric}

In this second approach, we forget about the discrete atomic distribution 
and consider instead the   
cloud as an effective 
homogeneous medium with a dielectric constant 
$\epsilon_{\rm eff}(\omega)=\epsilon_{\rm eff}'(\omega)+i\epsilon_{\rm eff}''(\omega)$.
In this {\it macroscopic} description the absorbed power
is given by the imaginary part of 
the dielectric constant and the macroscopic, 
ensemble-averaged field $\langle {\bf E}({\bf r})\rangle$ 
inside the medium~\cite{Jackson}:
\begin{eqnarray}\label{Eq:Pabs_macro}
P_{\rm abs}^{\rm macro}&=&\int_V{1\over 2}{\rm Re}[\langle {\bf J}\rangle\cdot\langle {\bf E}^*\rangle]\, dV\\\nonumber
&=&{\omega\over 2}\epsilon_0\epsilon_{\rm eff}''(\omega)\int_V|\langle {\bf E}({\bf r})\rangle|^2\, dV\ ,
\end{eqnarray} 
with $\langle {\bf J}\rangle=-i\omega \epsilon_0 [\epsilon_{\rm eff}(\omega)-1]\langle{\bf E}\rangle$.
A detailed balance derived in  Appendix~\ref{Appendix:Detailed_balance} shows  
that the total power $P_{\rm ext}$ taken from the incident field  
is the sum of the power absorbed by the cloud 
and of the scattered power $P_{\rm sc}^{\rm macro}$, i.e., 
$P_{\rm ext}=P_{\rm abs}^{\rm macro}+P_{\rm sc}^{\rm macro}$. 
Now, in a {\it microscopic} description, this same  
power $P_{\rm ext}$
is also the sum of the coherent and incoherent scattered 
powers, defined after ensemble averaging by Eq.~(\ref{Eq:Pcoh}) and (\ref{Eq:Pinc}): 
$P_{\rm ext}=P_{\text{coh}}+P_{\text{incoh}}$
\footnote{We have assumed 
here that the atoms are elastic scatterers 
that do not absorb light as e.g. particles in China ink would.}. 
Finally, the coherent scattered power defined by Eq.~(\ref{Eq:Pcoh})
is the power scattered in the macroscopic approach, 
$P_{\text{coh}}=P_{\rm sc}^{\text{macro}}$, leading to 
the identification of the absorbed power $P_{\rm abs}^{\rm macro}$ 
with the power $P_{\rm incoh}$ incoherently scattered by the cloud, i.e.:
\beq\label{Eq:Link_coh_incoh}
\frac{\epsilon_0c}{2}\oint_\Sigma\langle|\delta\textbf{E}_{\rm sc}|^2\rangle\, dS=
{\omega\over 2}\epsilon_0\epsilon_{\rm eff}''(\omega)\int_V|\langle {\bf E}({\bf r})\rangle|^2\, dV\ .
\eeq
As a consequence,
the homogenization criteria $P_{\rm incoh}=0$ requires that the 
imaginary part of the dielectric constant is negligible \footnote{Care 
must be taken that the fact that we can describe 
the incoherent power by the imaginary part of the effective (i.e. ensemble-averaged) 
dielectric constant does not imply that the incoherent field 
itself is described by an effective dielectric constant.}. 

We can now discuss the behavior of the coherent and incoherent scattered powers
as a function of the detuning shown in Fig.~\ref{Fig:spectrum_resonance}. 
As we studied in~\cite{Nick2016}, once we know the effective dielectric constant $\epsilon_{\rm eff}(\omega)$,
we can calculate the electric field inside the cloud considered as a continuous medium. We find 
resonances for certain values of the  frequency $\omega$, precisely 
corresponding to the polaritonic modes. When hitting a resonance, the 
electric field $\langle {\bf E}\rangle$ inside the medium 
is large. Consequently the scattered field, and therefore the coherent power, is large as well.
The variation of the coherent scattered power  shown in Fig.~\ref{Fig:spectrum_resonance} thus 
reflects the mode structure of the effective homogeneous particle 
equivalent to the cloud. We have for example checked
numerically that the frequency corresponding to the maximum 
of the broad peak on the red side of the resonance is 
proportional to the length of the cloud, a signature of a shape resonance in an object of finite size. 

The fact that the incoherent and coherent scattered powers  
show similar behavior as a function of $\omega$
is a direct consequence of 
Eq.~(\ref{Eq:Link_coh_incoh}): if $\epsilon_{\rm eff}''(\omega)$ were independent of $\omega$
the two powers would vary identically with $\omega$. As we showed in Ref.~\cite{Nick2016},
 $\epsilon_{\rm eff}''(\omega)$ is in fact a broad, resonant function and the incoherent scattered power 
 combines the shape resonance of $\langle\bf E\rangle$ with the resonant behavior of $\epsilon_{\rm eff}''(\omega)$. 

Finally, we can also understand why the coherent and incoherent scattered powers 
saturate when the number of atoms increases. 
For large atom numbers, the cloud behaves like a sharp
object.
The coherent scattering corresponds to the diffraction pattern of 
this object~\cite{Zangwill2012,Bohren}, and is 
only dependent on its shape~\footnote{As an example, the scattering cross section 
of an homogeneous particle with a size larger than the wavelength of the scattered 
light is twice the geometrical cross section 
perpendicular to the incoming wave direction, 
and is therefore independent of
the number of atoms in the particle.}.  
To understand why the incoherent power also saturates at large atom numbers, 
we rely on the expression of  
the coherent  power scattered by an object of volume 
$V$ and dielectric constant $\epsilon_{\rm eff}(\omega)$~\cite{Zangwill2012}:
\begin{eqnarray}
P_{\rm coh}&=&{\epsilon_0\omega^4\over 32\pi^2 c^3}\, |\epsilon_{\rm eff}(\omega)-1|^2 \\\nonumber
&& \left|\int_{4\pi} d\Omega\, {\bf e}_r\times\int_V d^3{\bf r}' \langle {\bf E}({\bf r}')\rangle\, e^{-i{\bf k}\cdot{\bf r}'}\right|^2\ ,
\end{eqnarray}
with ${\bf e}_r$  the unit vector in a given scattering direction. 
If the coherent scattered power saturates for large atom numbers, 
as it should based on the 
diffraction argument above, this formula indicates that the 
coherent field inside the object and the dielectric constant must saturate. 
Therefore, according to Eq.~(\ref{Eq:Link_coh_incoh}) the incoherent power must saturate as well.
 
\section{Influence of the correlations in the positions of the atoms}\label{Sec:correlatio_positions}

\begin{figure}
\includegraphics[width=\linewidth]{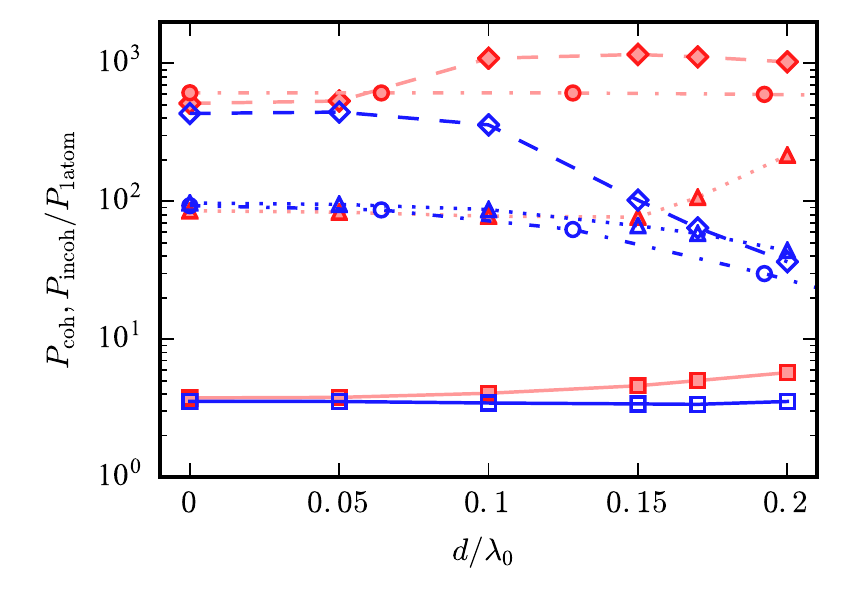}
\caption{Influence of position correlations on  incoherent (blue open symbols) and coherent 
(red plain symbols) scattering calculated for $N=100$ as 
a function of the diameter $d$ of the spherical exclusion volume around each 
atom, for various detunings $\Delta$.
Circles: $\Delta = -10^4\Gamma_0$, 
diamonds: $\Delta = -5\Gamma_0$, 
triangles: $\Delta = -2\Gamma_0$.
and squares: $\Delta = 0$.}
\label{Fig:correlation_positions}
\end{figure}

The impossibility to reach homogenization for an ensemble of cold atoms 
presented in Fig.~\ref{Fig:faroff_to_near_resonance} makes us wonder if there exists any
resonant system, which can reach the homogenization regime. 
In this Section, we show that 
a way to make this ensemble of scatterers 
homogenized  
consists in introducing spatial correlations 
in the positions of the atoms in order to reduce the fluctuations 
of the spatial distribution~\cite{Frisch1968,Lagendijk1996,Greffet2007}. 
We are guided here by the well-known fact that a pure liquid scatters  
less light than a gas, although being much denser:  the presence of 
position correlations in the liquid inhibits incoherent scattering. 
Another situation is the transparency of the cornea, which is 
due to spatial correlations~\cite{Hart1969}. 

Here we implement spatial correlations by introducing a spherical exclusion volume with diameter $d$ 
around each scatterer~\footnote{In this 
calculation, the number of atoms $N$ is kept constant. 
If for a given distribution we find that two atoms are distant by less than $d$, 
we draw a new configuration with the same number of atoms.}. 
This diameter sets the minimum distance between nearest-neighbor scatterers.
Figure~\ref{Fig:correlation_positions} presents the coherent and incoherent
scattered powers as a function of the diameter $d$ of the exclusion volume
for $N=100$ and various detunings. We observe that for $\Delta$ 
ranging from $-10^4\Gamma_0$ to $\sim-10\Gamma_0$
incoherent light scattering is
reduced by introducing spatial order in the system, while the coherent scattering
is weakly  affected. As a consequence, this procedure leads to homogenization 
according to the stringent criterion $P_{\rm incoh}= 0$. Qualitatively, this suppression
comes from the fact that the exclusion volume thwarts the formation of dimer modes. 

Closer to resonance, the behavior is more complex. 
For detunings in the range $-5\le\Delta/\Gamma_0<0$, we observe an  increase of 
the coherent power, while the incoherent 
decreases. At resonance ($\Delta=0$), 
coherent power still increases with $d$, 
while the incoherent power remains approximatively constant. 
These findings seem to indicate 
that the spatial correlations lead to at least the weak homogenization criterion, with 
a decrease of the ratio $P_{\rm incoh}/P_{\rm coh}$.
We could not explore  exclusion diameters larger than $0.2\lambda_0$, while maintaining the number 
of atoms constant~\footnote{When increasing the diameter $d$, the probability to find 
an exclusion volume with only one atom becomes too small. }, and therefore could not check 
the point when the stringent criterion starts to be valid. 


\section{Effect of non-radiative losses on near-resonance light scattering}\label{Sec:NRlosses}

In this last Section, we discuss the influence of non-radiative losses on 
light scattering. We know that, for example, a suspension of non-resonant 
absorbing particles, such as  a droplet of China ink consisting 
of a suspension of colloidal carbon nanoparticles, does not scatter light if the absorption 
cross section is much larger than the scattering cross section of each particle.  
In this case, all the energy gets absorbed by the particles, converted into heat, and 
scattering can be neglected. 
Guided by this example of a non-resonant situation, we introduce here non-radiative 
losses and study if homogenization according to the strong criterion 
($P_{\rm incoh} = 0$) is reached in the resonant case as well. 
Although this procedure would not apply to a cold atomic ensemble, it would be relevant 
for ensembles of e.g. molecules or quantum dots coupled to phonons. 

\begin{figure}
\includegraphics[width=1\linewidth]{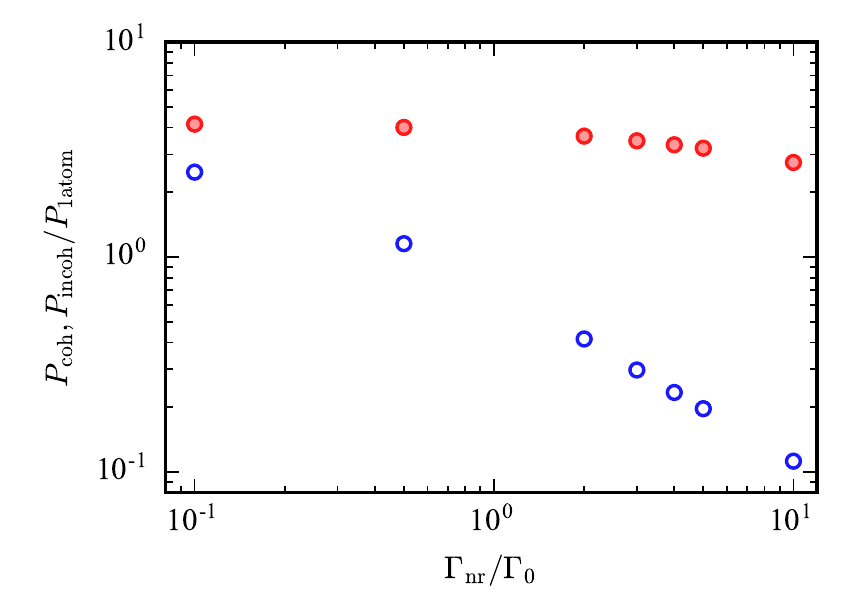}
\caption{Coherent (red plain circles) and incoherent (blue open circles) scattered powers on resonance ($\Delta =0$) for $N=450$ atoms as a function of the non-radiative decay rate $\Gamma_{\rm nr}$. All powers are normalized by the power scattered by a single atom at resonance.}
\label{Fig5_nonradiative}
\end{figure}

We report in Fig.~\ref{Fig5_nonradiative} the coherent and incoherent powers 
calculated for resonant light scattering for different values of the ratio 
of the non-radiative loss rate $\Gamma_{\rm nr}$ relative to the radiative 
loss rate $\Gamma_0$. 
We observe that as the amount of non-radiative losses increases 
the incoherent power gets significantly reduced, whereas the coherent 
power is only slightly affected. 

This observation can be understood by using the collective mode picture.
By introducing non-radiative losses characterized by the rate $\Gamma_{\rm nr}$, 
the only change to the (complex) eigenfrequency of a 
collective eigenmode $\beta$ is to replace its value  
$\tilde{\omega}_\beta=\omega_0+\Omega_\beta
-i\frac{\Gamma_{\beta}}{2}$ by $\tilde{\omega}_\beta-i\frac{\Gamma_\text{nr}}{2}$. 
As a consequence, the modes with radiative decay rates 
$\Gamma_\beta<\Gamma_{\rm nr }$
damp in a time $1/\Gamma_{\rm nr}$, irrespective of their radiative damping rate. 
Therefore, when  $\Gamma_{\rm nr}>\Gamma_0$, 
the sub-radiant modes ($\Gamma_\beta<\Gamma_0$) 
decay non-radiatively  very quickly and therefore hardly 
contribute to the scattering in steady state. As they are in particular 
responsible for the incoherent 
scattering, this one is suppressed.  
By contrast, light scattered by a polaritonic (super-radiant) mode is not 
affected by the non-radiative decay as long as
$\Gamma_\beta>\Gamma_{\rm nr }>\Gamma_0$. 
These polaritonic modes lead to  coherent scattering.  
Finally, when the radiative decay rate of the most super-radiant mode gets smaller 
than $\Gamma_{\rm nr}$, even the coherent scattering is suppressed.
Therefore, the introduction of non-radiative losses appears 
as an efficient way to achieve homogenization 
in the sense of suppressing incoherent light scattering. 

\section{Conclusion}

As a conclusion, we have studied theoretically the concept of 
homogenization in optics using an ensemble of resonant 
scatterers dense enough for the usual condition for 
homogenization, viz. $\rho\lambda^3 \gg 1$, to be reached. 
We have introduced two criteria to define the homogenization 
regime in terms of incoherent and coherent scattered powers. 
When the excitation field is tuned very far from the resonance of 
the scatterers, we recovered the well-known scaling laws as a function 
of the atom number $N$, leading to a suppression of the ratio 
$P_{\rm incoh}/P_{\rm coh}$
at large $N$. 
However, when excited in a broad frequency range around  
the resonance, we observed that 
none of the criteria for homogenization apply, meaning 
that the condition $\rho\lambda^3\gg 1$ is not sufficient to 
characterize the homogenized regime. We interpreted 
this result as an effect of the dipole-dipole interactions 
between the atoms, which implies a description of scattering in terms of 
collective modes rather than as a sequence of individual scattering events. 
Finally, we showed that, although 
homogenization can never be reached for a dense
ensemble of randomly positioned laser-cooled atoms,  it becomes possible if one introduces 
spatial correlations in the positions of the atoms 
or non-radiative losses, such as would be the case for organic molecules 
or quantum dots coupled to a phonon bath.
 
\begin{acknowledgements}
We thank R.~Carminati for fruitful discussions. We acknowledge support from the French National Research Agency (ANR) as part of the ``Investissements d'Avenir'' program (Labex PALM, ANR-10-LABX-0039), and R\'egion Ile-de-France (LISCOLEM project). N.J.~S. is supported by Triangle de la Physique and Universit\'e Paris-Sud. J.-J.~G. acknowledges support from Institut Universitaire de France and the SAFRAN-IOGS chair on Ultimate Photonics.
\end{acknowledgements}

\begin{appendix}

\section{Scattering by an ensemble of atoms in the single-scattering limit}\label{Appendix:single_scattering}

In this Appendix, we summarize textbook arguments about 
scattering from ensemble of scatterers in the single-scattering limit, 
valid when the mean-free path
$\ell_\text{sc}=1/[\rho\sigma_\text{sc}(\omega)]$ is much larger than the size of the ensemble.
The power scattered at a distance $r$ in the direction ${\bf k}$ by the ensemble 
of $N$ atoms placed at positions ${\rm r}_j$ is then proportional to
\beq
|{\bf E}_{\rm sc}({\bf k})|^2 = |{\bf E}_{0,\rm sc}({\bf k})|^2\, S({\bf k}-{\bf k}_{\rm L})\ ,
\eeq
where ${\bf k}_{\rm L}$ is the wave vector of the incident plane wave,  
$ {\bf E}_{0,\rm sc}({\bf k})$ the field scattered by a single atom, and 
\beq
S({\bf q})=|\sum_{j=1}^N e^{i{\bf q}\cdot {\bf r}_j}|^2 
\end {equation}
the structure factor of the cloud. The ensemble-average power in the direction 
${\bf k}={\bf k}_{\rm L} +{\bf q}$ is then proportional to $\langle S({\bf q})\rangle$
and thus:
\beq\label{Eq:scattering_coh_incoh}
\langle |{\bf E}_{\rm sc}({\bf k})|^2 \rangle\ \propto N(1-\langle e^{i{\bf q}\cdot({\bf r}-{\bf r}')}\rangle)+ N^2\langle e^{i{\bf q}\cdot({\bf r}-{\bf r}')}\rangle\ .
\eeq
Here, the average phase factor s
\beq
\langle e^{i{\bf q}\cdot({\bf r}-{\bf r}')}\rangle = \int_Vd^3{\bf r}d^3{\bf r}' P({\bf r},{\bf r}')e^{i{\bf q}\cdot({\bf r}-{\bf r}')}\ , 
\eeq
with $P({\bf r},{\bf r}')$ the joint probability distribution to find a particle at position ${\bf r}$ and
${\bf r}'$.
When the positions are not correlated, $P({\bf r},{\bf r}')= \rho({\bf r})\rho({\bf r}')/N^2$, 
with $\rho({\bf r})$ the spatial density distribution, and 
the term $\langle e^{i{\bf q}\cdot({\bf r}-{\bf r}')}\rangle$ 
is the diffraction pattern of the cloud:
\beq
\langle e^{i{\bf q}\cdot({\bf r}-{\bf r}')}\rangle =
\left| {1\over N}\int_Vd^3{\bf r} \rho({\bf r}) e^{i{\bf q}\cdot{\bf r}}\right|^2\ . 
\eeq
The $N^2$ term in Eq.~(\ref{Eq:scattering_coh_incoh}) is  the coherent component, 
and dominates in the solid angle 
corresponding to diffraction. In the other directions, the power 
is  proportional to $N$ and corresponds
to the incoherent scattering. It is almost isotropic. 

\section{Contribution of the dimer modes to incoherent scattering in the intermediate detuning regime}\label{Appendix:dimer_modes}

Here we derive a semi-classical model to calculate the critical number of atoms $N_{\rm c}$ 
where the transition between scattering by individual atoms to scattering by dimers occurs (Sec.~\ref{Sec:Interpret_modes}). This model is inspired by the one used to calculate the rate of
light-assisted collisions in cold atomic samples~\cite{GallagherPritchard89,WeinerRMP}.

As explained in Sec.~\ref{Sec:Interpret_modes}, in the intermediate detuning regime
the  incoherent scattering  is due to the super-radiant 
pairs of atoms. We first note that when two atoms are separated by $r\ll 1/k$, 
their interaction energy is dominated by the dipolar (near-field) term: 
$U(r)\sim \hbar\Gamma_0/(kr)^3$. 
To estimate the incoherent scattered
power, we start by calculating the number of excited pairs as follows: 
for a given detuning $|\Delta|$, the light is resonant with the excitation 
of a pair of atoms located at a relative distance $r_{\rm ex}$
such that 
$|\Delta|\sim \Gamma_0/(kr_{\rm ex})^3$. 
The number of excited dimers is 
then the product of the atom number $N$ with 
the number of atoms in a shell
of radius $r_{\rm ex}$, $(N/V) 4\pi r_{\rm ex}^2\Delta r$, with $V$ the cloud volume, 
$N$ the atom number and $\Delta r$ the thickness of the shell. 
The latter one is determined by writing that the collective width $2\Gamma_0$
of a super-radiant pair is $U'(r_{\rm ex})\Delta r/\hbar = 3 \Gamma_0/(k r_{\rm ex})^3 (\Delta r / r_{\rm ex})$. As a consequence, assuming that approximatively 
half the dimers are super-radiant, 
the number of super-radiant pairs is $N_{\rm p} \sim (N^2/2)[8\pi/(3k^3V)](\Gamma_0/\Delta)^2$.
The power scattered by a super-radiant pair (decay rate $2\Gamma_0$, saturation intensity of the dimer transition $2I_{\rm sat}$, with $I_{\rm sat}$ the atomic saturation intensity) 
irradiated by a light with intensity $I\ll I_{\rm sat}$ 
is $P_{\rm dimer}=\hbar\omega_0 (2\Gamma_0) I/ (4I_{\rm sat})$. 
Finally, the power scattered by a single atom
irradiated by the same light  detuned 
by $|\Delta|$ with respect to the atomic transition 
is $P_0\approx\hbar\omega_0\Gamma_0 I/(2I_{\rm sat})(\Gamma_0/2\Delta)^2$. 
Combining the expressions above, we find the ratio of  
the total incoherent  power scattered by the cloud to the power scattered by a single atom: 
$P_{\rm incoh}/P_0 \sim N^2 [16\pi / (3k^3 V)] $. 
The critical number of atoms $N_{\rm c}$ 
where the transition between scattering by individual atoms to scattering by dimers occurs
is thus $N_c \sim 3k^3 V/ (16 \pi)$. The transition point 
is thus predicted to be independent of the detuning, 
in agreement with numerical calculations performed 
for various detunings.  
Also, for our parameters, $N_c\approx 25$, in good  agreement with the value observed
in Fig.~\ref{Fig:faroff_to_near_resonance}, considering the simplicity of the model.

We now calculate the coherent power $P_{\rm coh}^{\rm dim}$ 
scattered by the dimers and compare
it to the coherent scattering due to individual atoms $P_{\rm coh}^{\rm at}$. This latter one 
is given by $P_{\rm coh}^{\rm at} =  N^2 P_0 \, \Omega/(4\pi)$, as for the intermediate 
detunings the cloud is still in the single scattering  regime. 
Here, $\Omega/(4\pi)$
is the solid angle of the 
diffraction pattern. 
The power coherently scattered by the dimers is  $P_{\rm coh}^{\rm dim}= N_{\rm p}^2P_{\rm dimer}\, \Omega/(4\pi)$
as the dimers are spread in a dilute way in the same volume than the atoms.
Finally, we get $P_{\rm coh}^{\rm dim}/ P_{\rm coh}^{\rm at}= [8\pi/(3k^3V)]^2 N^2(\Gamma_0/\Delta)^2$. 
For $N\lesssim 500$ and $|\Delta|/\Gamma_0=100-500$,  $P_{\rm coh}^{\rm dim}/ P_{\rm coh}^{\rm at}\ll1$, thus making the 
contribution of the dimers to coherent scattering negligible.

\section{Derivation of the detailed balance used in Sec.\ref{Sec:interpret_dielectric}}\label{Appendix:Detailed_balance}

Here we give an explicit derivation of Eq.~\ref{Eq:Link_coh_incoh}. 
We consider the ensemble (volume $V$) of discrete atoms 
and a spherical surface $\Sigma$ with a radius $r\gg 1/k$. 
The conservation of energy 
in steady-state state tells us that the flux of the time-averaged Poyting vector 
through $\Sigma$ compensates for the 
time-averaged power dissipated by the current in the volume $V$:
\beq
{1\over 2}{\rm Re}\Big[\oint_{\Sigma} {\bf E}\times {\bf H}^*\cdot  d{\bf S} +\int_V{\bf J}\cdot{\bf E}^*\, dV\Big]=0 \ , 
\eeq
with ${\bf H}={\bf B}/\mu_0$ (non magnetic medium) and $\bf J$ the 
current density in the ensemble.  This 
expression is valid in a microscopic model. 
We now decompose the fields into the incoming component, 
the coherent scattered component and the incoherent part: 
${\bf E}={\bf E}_{\rm i}+\langle{\bf E}_{\rm sc}\rangle+\delta{\bf E}_{\rm sc} $ and 
${\bf H}={\bf H}_{\rm i}+\langle{\bf H}_{\rm sc}\rangle+\delta{\bf H}_{\rm sc} $, 
with $\langle.\rangle$
denoting an ensemble average. We also decompose
the current density into an ensemble average and a fluctuating 
part: ${\bf J}=\langle{\bf J}\rangle+\delta{\bf J} $.  
As $\oint_{\Sigma} {\bf E}_{\rm i}\times {\bf H}_{\rm i}^*\cdot  d{\bf S} =0$,  
the extinction power 
$P_{\rm ext}$ taken from the incident field is, 
after expansion and ensemble  average:
\begin{eqnarray}\label{Eq:Balanced1}
&&P_{\rm ext}=-{1\over 2}{\rm Re}\Big[\oint_{\Sigma} ({\bf E}_{\rm i}\times \langle{\bf H}_{\rm sc}^*\rangle+
\langle{\bf E}_{\rm sc}\rangle\times {\bf H}_{\rm i}^*)\cdot d{\bf S}\Big]\nonumber
\\
&=&-{1\over 2}{\rm Re}\Big[\oint_{\Sigma}\langle {\bf E}_{\rm sc}\rangle\times \langle{\bf H}_{\rm sc}^*\rangle\cdot d{\bf S}+
\oint_{\Sigma}\langle \delta{\bf E}_{\rm sc}\times \delta{\bf H}_{\rm sc}^*\rangle\cdot d{\bf S}\nonumber
\\
&&+\int_V  \langle{\bf J}\rangle\cdot \langle{\bf E}^*\rangle\, dV+\int_V  \langle\delta{\bf J}\cdot \delta{\bf E}^*\rangle\, dV\Big]\ ,
\end{eqnarray}
with $\langle{\bf E}\rangle= {\bf E}_{\rm i}+\langle{\bf E}_{\rm sc}\rangle$. 
For a cloud of cold atoms, no energy is dissipated in the medium as all incident power is re-scattered. 
Therefore:
\beq
\int_V  \langle {\bf J}\cdot {\bf E}^*\rangle\, dV=\int_V  \langle\delta{\bf J}\cdot \delta{\bf E}^*\rangle\, dV+\int_V  \langle{\bf J}\rangle\cdot \langle{\bf E}^*\rangle\, dV=0\ .
\eeq
In this case, $P_{\rm ext}=P_{\rm coh}+P_{\rm incoh}$ using the definitions of the 
coherent (Eq.~\ref{Eq:Pcoh}) and incoherent (Eq.~\ref{Eq:Pinc})
scattered powers.

Now, for the homogeneous medium with an effective permittivity 
equivalent to the ensemble of atoms, $\delta{\bf J}={\bf 0}$, 
$\delta{\bf E}=\delta{\bf E}_{\rm sc}={\bf 0}$ and $\delta{\bf H}_{\rm sc}={\bf 0}$, and Eq.~(\ref{Eq:Balanced1}) yields
\begin{eqnarray}
&&P_{\rm ext}=-{1\over 2}{\rm Re}\Big[\oint_{\Sigma} 
({\bf E}_{\rm i}\times \langle{\bf H}_{\rm sc}^*\rangle+
\langle{\bf E}_{\rm sc}\rangle\times {\bf H}_{\rm i}^*)\cdot d{\bf S}\Big]\nonumber\\
&=&{1\over 2}{\rm Re}\Big[\oint_{\Sigma}\langle {\bf E}_{\rm sc}\rangle\times \langle{\bf H}_{\rm sc}^*\rangle\cdot d{\bf S}+
\int_V  \langle{\bf J}\rangle\cdot \langle{\bf E}^*\rangle\, dV\Big]\ .
\end{eqnarray}
As $P_{\rm coh}={1\over 2}{\rm Re}\Big[\oint_{\Sigma}\langle {\bf E}_{\rm sc}\rangle\times \langle{\bf H}_{\rm sc}^*\rangle\cdot d{\bf S} \Big]$, we get by identification:
\beq
P_{\rm incoh}= {1\over 2}{\rm Re}\Big[\int_V  \langle{\bf J}\rangle\cdot \langle{\bf E}^*\rangle\, dV\Big]\ , 
\eeq
or equivalently, $P_{\rm incoh}=P_{\rm abs}^{\rm macro}$\ .

\end{appendix}

\end{document}